\documentclass[11pt]{article}
\usepackage{amsmath}
\usepackage{amsfonts}
\usepackage{amssymb}
\usepackage[dvips]{epsfig}

 \advance \textwidth by -2in
 \advance \textheight by -3in
\def\##1{{\bf #1}}
\def\=#1{\underline{\underline{#1}}}

\def\+#1{\underline{\bf #1}}
\def\*#1{\underline{\underline{\bf #1}}}

\def\eps{\epsilon}
\def\epso{\epsilon_{\scriptscriptstyle 0}}
\def\muo{\mu_{\scriptscriptstyle 0}}
\def\ko{k_{\scriptscriptstyle 0}}

\def\.{\mbox{ \tiny{$^\bullet$} }}

\def\ux{\#{u}_x}
\def\uy{\#{u}_y}
\def\uz{\#{u}_z}

\def\le{\left(}
\def\ri{\right)}
\def\les{\left[}
\def\ris{\right]}
\def\lec{\left\{}
\def\ric{\right\}}

\def\l#1{\label{#1}}
\def\r#1{(\ref{#1})}

\def\eps{\epsilon}
\def\epso{\epsilon_0}
\def\muo{\mu_0}
\def\ko{k_0}

\def\.{\mbox{ \tiny{$^\bullet$} }}

\def\ux{\#{u}_x}
\def\uy{\#{u}_y}
\def\uz{\#{u}_z}

\def\le{\left(}
\def\ri{\right)}
\def\les{\left[}
\def\ris{\right]}
\def\lec{\left\{}
\def\ric{\right\}}

\def\c#1{\cite{#1}}
\def\l#1{\label{#1}}
\def\r#1{(\ref{#1})}

\begin{document}

\begin{center} {\bf {\LARGE Scattering loss in electro--optic particulate composite materials}}
\end{center} \vskip 0.2cm
 \vskip 0.2cm

\noindent  {\bf {\large Tom G. Mackay$^a$ and Akhlesh Lakhtakia$^b$}
} \vskip 0.4cm

\noindent {\sf $^a$School of Mathematics\\
\noindent James Clerk Maxwell Building\\
\noindent University of Edinburgh\\
\noindent Edinburgh EH9 3JZ, United Kingdom\\
email: T.Mackay@ed.ac.uk} \vskip 0.4cm

\noindent {\sf $^b$CATMAS~---~Computational \& Theoretical Materials Sciences Group \\
\noindent Department of Engineering Science \& Mechanics\\
\noindent 212 Earth \& Engineering Sciences Building\\
\noindent Pennsylvania State University, University Park, PA
16802--6812, USA\\
email: akhlesh@psu.edu} \vskip 2.4cm

\begin{center} {\bf Abstract}\end{center}
The effective permittivity dyadic of a composite material containing
particulate constituent  materials with one constituent having the
ability to display the Pockels effect is  computed, using an
extended version of the strong--permittivity--fluctuation theory
which takes account of both the distributional statistics of the
constituent particles and their sizes. Scattering loss, thereby
incorporated in the effective electromagnetic response of the
homogenized composite material, is significantly affected by the
application of a low--frequency (dc) electric field.

\vspace{8mm}


\section{Introduction}

Two (or more) materials, each composed of particles which are small
compared to all relevant wavelengths, may be blended together to
create an effectively homogeneous material. By judiciously
selecting the constituent materials, as well their relative
proportions, particle shapes, orientations and sizes, and
distributional statistics, the homogenized composite material (HCM) can be
made to display desirable magnitudes of effective constitutive parameters
 \c{L96, Tor}. Furthermore, an HCM can exhibit effective
constitutive parameters which are either not exhibited at all by
its constituent materials, or at least not exhibited to the same
extent by its constituent materials \c{M05,JKSL}. A prime example of such
\emph{metamaterials} is provided by HCMs which support
electromagnetic planewave
propagation with negative phase velocity \c{ML06}.

The focus of this paper is on the electromagnetic constitutive
properties of HCMs \c{L96,Neela}. If one (or more) of the
constituent materials exhibits the Pockels effect \c{Boyd_92}, then
a further degree of control over the electromagnetic response
properties of an HCM may be achieved. That is, post--fabrication
dynamical control of an HCM's performance may be achieved through
the application of a low--frequency (dc) electric field. For such an
HCM, the potential to engineer its electromagnetic response
properties (i) at the fabrication stage by selection of the
constituent materials and particulate geometry, and (ii) at the
post--fabrication stage by an applied field, is of considerable
technological importance in the area of smart materials \c{Hoff}.

The opportunities
 offered by the Pockels effect for tuning the response
 properties of composite materials
have recently been highlighted for  photonic
band--gap engineering \c{Lajp,Li} and
HCMs \c{LM07}. In particular, the predecessor study exploiting the well--known
Bruggeman homogenization formalism revealed that the greatest degree
of control over the HCM's constitutive parameters is achievable when
the constituent materials are distributed as oriented and highly aspherical particles
and have high electro--optic coefficients \c{LM07}.
However, the Bruggeman formalism may not take predict the scattering
loss in a composite material adequately enough, as it does not take
into account positional correlations between the particles.
Therefore, in the following
sections of this paper, we implement a more sophisticated homogenization approach
based on the strong--permittivity--fluctuation theory (SPFT) which
enables us to investigate the effect of the dc electric field on the degree of
scattering loss in electro--optic HCMs.

A note about notation:  Vectors are in boldface, dyadics are double
underlined.  The inverse, adjoint, determinant and trace of a dyadic
$\=M$ are represented as $\=M^{-1}$, $\=M_{\,adj}$,
$\mbox{det}\,\le \, \=M \, \ri $, and $\mbox{tr}\,\le \, \=M \, \ri
$, respectively. A Cartesian coordinate system with unit vectors
$\#u_{x,y,z}$ is adopted. The identity  dyadic is written as $\=I$,
and the null dyadic as $\=0$. An $\exp(-i\omega t)$ time--dependence
is implicit
 with $i = \sqrt{-1}$, $\omega$ as angular frequency, and $t$ as time.
  The permittivity and permeability   of free space are denoted by $\epso$ and $\muo$,
  respectively; and the free--space wavenumber is $\ko = \omega
  \sqrt{\epso \muo}$.

\section{Theory}

Let us now apply the  SPFT to estimate the effective permittivity dyadic of an HCM
arising from two particulate constituent materials, one of which
exhibits the Pockels effect. Unlike conventional approaches to
homogenization, as typified by the usual versions of the  much--used
Maxwell Garnett and Bruggeman formalisms \c{L96,Neela}, the SPFT can take
detailed account of the distributional statistics of the constituent
particles \c{TK81,Genchev,Z94}. The extended version of the SPFT
implemented here \emph{also} takes into account the sizes of the
constituent particles \c{M04}.

\subsection{Constituent materials}

The  two constituent materials are labeled $a$ and $b$. The
particles of both materials   are ellipsoidal, in general. For
simplicity, all constituent particles are taken to have the same
shape and orientation,
 as specified by the dyadic
\begin{equation}
 \=U^{} = \le \alpha_1 \alpha_2 \alpha_3 \ri^{-1/3} \sum_{K=1}^3\,
\alpha_K\,\#a_K\#a_K,
\end{equation}
 with $\alpha_K>0\,\forall K \in\les1,3\ris$ and the three unit
vectors $\#a_{1,2,3}$ being mutually orthogonal. Thus, the surface
of  each constituent particle may be parameterized by
\begin{equation}
\#r^e (\theta, \phi) = \eta \, \=U \. \hat{\#r} (\theta, \phi),
\end{equation}
where $\hat{\#r} (\theta, \phi)$ is the radial unit vector in the direction specified by the
spherical polar coordinates $\theta$ and $\phi$. The size parameter
$\eta > 0$ is  a measure of the average linear dimensions of the
particle. It is fundamental to the  process of homogenization that
$\eta$  is  much smaller than all electromagnetic wavelengths \c{L96}.
However,  $\eta$ need not be vanishingly small but just be electrically
small \c{SL93,PLS94}.

Let
$V_a$ and $V_b$ denote the disjoint regions which contain the
constituent materials $a$ and $b$, respectively.
The constituent particles are  randomly distributed.
 The distributional statistics  are described in terms of moments of the
characteristic functions
\begin{equation}
\Phi_{ \ell}(\#r) = \left\{ \begin{array}{ll} 1, & \qquad \#r \in
V_{\, \ell},\\ & \qquad \qquad \qquad \qquad \qquad \qquad (\ell=a,b) . \\
 0, & \qquad \#r \not\in V_{\, \ell}, \end{array} \right.
\end{equation}
 The volume fraction of constituent material $\ell$, namely $f_\ell$ , is given by
the first statistical moment of
 $\Phi_{\ell}$ ;
 i.e., $\langle \, \Phi_{\ell}(\#r) \, \rangle = f_\ell$; furthermore,
 $f_a + f_b = 1$.
For the second statistical moment of $\Phi_{\ell}$,
 we adopt the physically motivated form \c{TKN82}
\begin{equation}
\langle \, \Phi_\ell (\#r) \, \Phi_\ell (\#r')\,\rangle = \left\{
\begin{array}{lll}
\langle \, \Phi_\ell (\#r) \, \rangle \langle \Phi_\ell
(\#r')\,\rangle\,, & & \hspace{10mm}  | \, \=U^{-1}\. \le   \#r - \#r' \ri | > L ,\\ && \hspace{25mm} \\
\langle \, \Phi_\ell (\#r) \, \rangle \,, && \hspace{10mm}
 | \, \=U^{-1} \. \le  \#r -
\#r' \ri | \leq L.
\end{array}
\right.
 \l{cov}
\end{equation}
The  correlation length $L>0$ herein is required to be much smaller
than the electromagnetic wavelength(s), but larger than the particle
size parameter $\eta$. The particular form of the covariance
function has little influence on SPFT estimates of the  HCM
constitutive parameters, at least for physically plausible
covariance functions \c{MLW01b}.

Next we turn to the electromagnetic constitutive properties of the constituent
materials. Material $a$ is  simply an isotropic dielectric material
with  permittivity dyadic $\=\eps^{(a)} = \eps^{(a)} \=I\,$  in the
optical regime. In contrast, material $b$ is more complicated as it
displays the Pockels effect. Its linear electro--optic properties
are expressed through the inverse of its  permittivity dyadic in the
optical regime, which  is written  as \c{LM07}
\begin{eqnarray}
\nonumber \les\=\eps^{(b)}\ris^{-1}&=& \frac{1}{\epso} \Bigg\{
\sum_{K=1}^3\les\le1/\epsilon _{K}^{(b)}+s_j\ri \,\#u_K\#u_K\ris\\[5pt]
&+&s_4 \le\#u_2\#u_3 +\#u_3\#u_2\ri +s_5 \le\#u_1\#u_3
+\#u_3\#u_1\ri +s_6 \le\#u_1\#u_2 +\#u_2\#u_1\ri \Bigg\},
  \label{Poc}
\end{eqnarray}
where
\begin{equation}
\label{sJ} s_J= \sum_{K=1}^3 r_{JK}\,E_{K}^{dc}\,,\quad
J\in\les1,6\ris.
\end{equation}
The unit vectors
\begin{equation}
\left.\begin{array}{l} \#u_1=-(\ux\cos\phi_b+\uy\sin\phi_b)
\cos\theta_b+\uz\sin\theta_b\\[5pt]
\#u_2=\ux\sin\phi_b-\uy\cos\phi_b\\[5pt]
\#u_3=(\ux\cos\phi_b+\uy\sin\phi_b)\sin\theta_b+\uz\cos\theta_b
\end{array}\right\}\,,\quad \theta_b\in\les0,\pi\ris\,,\quad
\phi_b\in\les0,2\pi\ris\,,
\end{equation}
pertain to the crystallographic structure of the material. In
\r{Poc} and \r{sJ}, $E_{K}^{dc}=\#u_K\.\#E^{dc}$, $K\in\les1,3\ris$,
are the Cartesian components of the dc electric field; $\epsilon
_{1,2,3}^{(b)}$ are the principal relative permittivity scalars in
the optical regime; and $r_{JK}$, $J\in\les1,6\ris$ and
$K\in\les1,3\ris$, are the $18$ electro--optic coefficients in the
traditional contracted or abbreviated notation for representing
symmetric second--order tensors \c{Auld}.  Correct to the first
order in the components of the dc electric field, which is
commonplace in electro--optics  \c{Yariv_Yeh}, we get the linear
approximation \c{L06_JEOS}
\begin{eqnarray}
\nonumber \=\eps^{(b)} &\approx& \epso
\Bigg\{
\sum_{K=1}^3\les\epsilon _{K}^{(b)}\le
1-\epsilon _{K}^{(b)}s_K \ri\,\#u_K\#u_K\ris\\[5pt]
&-&\epsilon _{2}^{(b)}\epsilon _{3}^{(b)}\,s_4 \le\#u_2\#u_3
+\#u_3\#u_2\ri -\epsilon _{1}^{(b)}\epsilon _{3}^{(b)}\,s_5
\le\#u_1\#u_3 +\#u_3\#u_1\ri -\epsilon _{1}^{(b)}\epsilon
_{2}^{(b)}\,s_6 \le\#u_1\#u_2 +\#u_2\#u_1\ri \Bigg\} \label{PocEps}
\end{eqnarray}
from (\ref{Poc}), provided that
\begin{equation}
\label{restriction} \lec {\substack{\max \\ K\in\les1,3\ris}}
\,\lvert \eps_K^{(b)}\rvert\ric\,\, \lec{\substack{\max \\
J\in\les1,6\ris}}\,\lvert s_J\rvert \ric\ll 1\,.
\end{equation}
The constituent material $b$ can be isotropic, uniaxial, or biaxial,
depending on the relative values of $\epsilon_1^{(b)}$,
$\epsilon_2^{(b)}$, and $\epsilon_3^{(b)}$. Furthermore, this
material may belong to one of 20 crystallographic classes of point
group symmetry, in accordance with the relative values of the
electro--optic coefficients.

In order to highlight the degree of electrical control which can be achieved over the HCM's
permittivity dyadic, we consider scenarios
wherein the influence of the Pockels effect is most conspicuous. Therefore,  the crystallographic and geometric
orientations of the constituent particles  are taken to be aligned from here onwards;
furthermore, $\#E^{dc}$ is aligned with the major
crystallographic/geometric principal axis. For convenience, the
principal crystallographic/geometric axes are taken to coincide with
the Cartesian basis vectors $\#u_{x,y,z}$.

\subsection{Homogenized composite material}

The bilocally approximated SPFT estimate of the
  permittivity dyadic of the HCM turns out to be \c{M04}
\begin{equation}
\=\eps^{HCM}_{}  = \epso \les \, \mbox{diag} \le \eps^{HCM}_x,
\,\eps^{HCM}_y, \,\eps^{HCM}_z \ri \ris =
 \=\eps^{cm} - \le\,\=I +
\=\Sigma \. \=D \,\ri^{-1}\.\=\Sigma\,. \l{SPFT_HCM}
\end{equation}
Herein, $ \=\eps^{cm}$ is the  permittivity dyadic of a homogeneous
comparison medium, which is delivered by solving the Bruggeman
equation \c{LM07}
\begin{eqnarray}
\nonumber &&f_a\, \=\chi^{(a)} + f_b \,\=\chi^{(b)} =\=0, \label{Brug}
\end{eqnarray}
where
\begin{equation}
\=\chi^{(\ell)} = \le\,\=\eps^{(\ell)} - \=\eps^{cm}\,\ri\.\les\,
\=I +  \=D\.\le\, \=\eps^{(\ell)} - \=\eps^{cm}\,\ri\,\ris^{-1},
\hspace{20mm} (\ell =a,b) \l{X_def}
 \end{equation}
are  polarizability density dyadics. The depolarization dyadic is
represented by the sum
\begin{equation}
\=D = \=D^{ 0} + \=D^{ > 0} (\eta).
\end{equation}
The depolarization contribution arising from vanishingly small
particle regions (i.e.,  $\eta \rightarrow 0$) is given by
\c{LM07,Michel}
\begin{equation}
\=D^{0} =
 \frac{1}{  4 \pi } \, \int^{2\pi}_{0} \; d\phi_q
\, \int^\pi_0\;  d\theta_q\;  \frac{\sin \theta}{\mbox{tr} \le
\,\=\eps^{cm}\.\=A \,\ri} \, \=A , \l{depol}
\end{equation}
 where
\begin{equation}
\=A = \mbox{diag} \,\les  \le \frac{\sin \theta_q \, \cos
\phi_q}{\#u_x \. \=U \. \#u_x}\ri^2,\, \le \frac{\sin \theta_q \,
\sin \phi_q}{\#u_y \. \=U \. \#u_y}\ri^2,\,\le \frac{\cos \theta_q
}{\#u_z \. \=U \. \#u_z}\ri^2\ris,
\end{equation}
 while the depolarization contribution arising from
particle regions of finite size (i.e., $\eta > 0$) is given by \c{M04}
\begin{eqnarray}
 \=D^{> 0} (\eta)   &=& \frac{\eta^3}{4 \pi} \int^{2 \pi}_{0} d
\phi_q \int^{\pi}_0
 d \theta_q\;\;
\frac{\sin \theta}{3 \, \Delta} \Bigg\{ \,  \les \,
 \frac{3 \le \kappa_+ -
\kappa_-  \ri}{2 \eta} + i \le \kappa^{\frac{3}{2}}_+  -
\kappa^{\frac{3}{2}}_-  \ri \ris
  \=\alpha  \nonumber \\ && +
i  \omega^2 \muo \,\le \kappa^{\frac{1}{2}}_+ -
\kappa^{\frac{1}{2}}_-  \ri  \=\beta \, \Bigg\}, \l{W_def}
\end{eqnarray}
with
\begin{eqnarray}
&&  \=\alpha =
  \les 2 \,\=\eps^{cm} - \mbox{tr} \le \, \=\eps^{cm} \, \ri
\, \=I \, \ris\. \=A - \mbox{tr} \le \,\=\eps^{cm}\.\=A\,\ri \,
\=I\,
 -  \, \frac{  \mbox{tr} \le \, \=\eps^{cm}_{\,adj}\.\=A\,\ri -
\les \, \mbox{tr} \le \, \=\eps^{cm}_{\,adj} \, \ri \, \mbox{tr} \le
\, \=A \, \ri \, \ris }{
 \mbox{tr} \le \, \=\eps^{cm}\. \=A \, \ri } \,  \=A, \, \nonumber \\
&&
\\
&& \=\beta =
 \=\eps^{cm}_{\,adj} -  \frac{  \det \le \, \=\eps^{cm} \, \ri }
{ \mbox{tr} \le \, \=\eps^{cm}\. \=A \, \ri} \, \=A \,,
\\
&&  \Delta =   \lec \les \mbox{tr} \le \,
\=\eps^{cm}_{\,adj}\.\=A\,\ri -
 \mbox{tr} \le \, \=\eps^{cm}_{\,adj}\, \ri \, \mbox{tr} \le
\, \=A \, \ri \, \ris^2 - 4 \det \le \, \=\eps^{cm} \, \ri \mbox{tr}
\le \, \=A \, \ri \,
 \mbox{tr} \le \, \=\eps^{cm}\. \=A \, \ri \ric^{\frac{1}{2}}, \\
&& \kappa_{\pm}  = \muo \omega^2 \frac{ \les \, \mbox{tr} \le \,
\=\eps^{cm}_{\,adj} \, \ri \, \mbox{tr} \le \, \=A \, \ri \, \ris -
\mbox{tr} \le \, \=\eps^{cm}_{\,adj}\.\=A\,\ri \pm \Delta}{2 \,
\mbox{tr} \le \, \=A \, \ri \,
 \mbox{tr} \le \, \=\eps^{cm}\. \=A \, \ri}.
\end{eqnarray}
The mass operator dyadic in \r{SPFT_HCM} is defined as
\begin{equation}
\=\Sigma = f_a f_b  \le\,\=\chi^{(a)} -
\=\chi^{(b)}\,\ri\.\=D^{>0}(L)\.\le\,\=\chi^{(a)} -
\=\chi^{(b)}\,\ri.
\end{equation}

\section{Numerical results and conclusion} \l{num_studies}

The prospects for electrical control over the HCM's permittivity
dyadic are explored by means of an illustrative example. In keeping
with the predecessor study based on the Bruggeman homogenization
formalism \c{LM07}, let us choose an example in which the influence
of the Pockels effect is highly noticeable. Thus, we set the
relative permittivity scalar $\eps^{(a)} = 1$, whereas the
constituent material $b$ has the constitutive properties of
potassium niobate \c{ZSB}:  $\epsilon_1^{(b)} = 4.72$,
$\epsilon_2^{(b)}= 5.20$, $\epsilon_3^{(b)}=5.43$, $r_{13}=34\times
10^{-12}$~m~V$^{-1}$, $r_{23}=6\times 10^{-12}$~m~V$^{-1}$,
$r_{33}=63.4\times 10^{-12}$~m~V$^{-1}$, $r_{42}=450\times
10^{-12}$~m~V$^{-1}$, $r_{51}=120\times 10^{-12}$~m~V$^{-1}$, and
all other $r_{JK}\equiv0$.  For all calculations, the volume
fraction is fixed at $f_a = 0.5$; the shape parameters are $\alpha_1
= \alpha_2 =1$ and $\alpha_3 = 9$ (so that the particles are prolate
spheroids); the crystallographic angles $\theta_b = \phi_b = 0$;
$E^{dc}_{1,2} = 0$ and the range for $E^{dc}_3 $ is dictated by
\r{restriction}, thereby allowing direct comparison of the SPFT
results with results for the Bruggeman formalism \c{LM07}. The SPFT
calculations were carried out for an angular frequency $\omega = \pi
\times 10^{15} \, \mbox{rad}$~$\mbox{s}^{-1}$, corresponding to a
free--space wavelength of 600 nm.

\begin{figure}[!ht]
\centering \psfull \epsfig{file=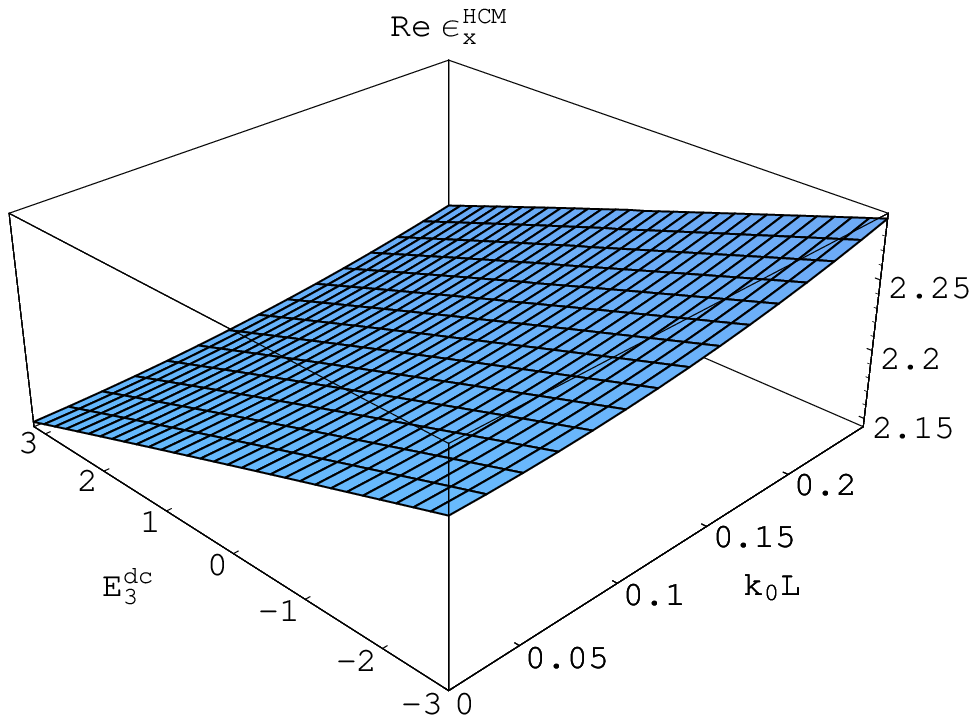,width=3.0in} \hfill
\epsfig{file=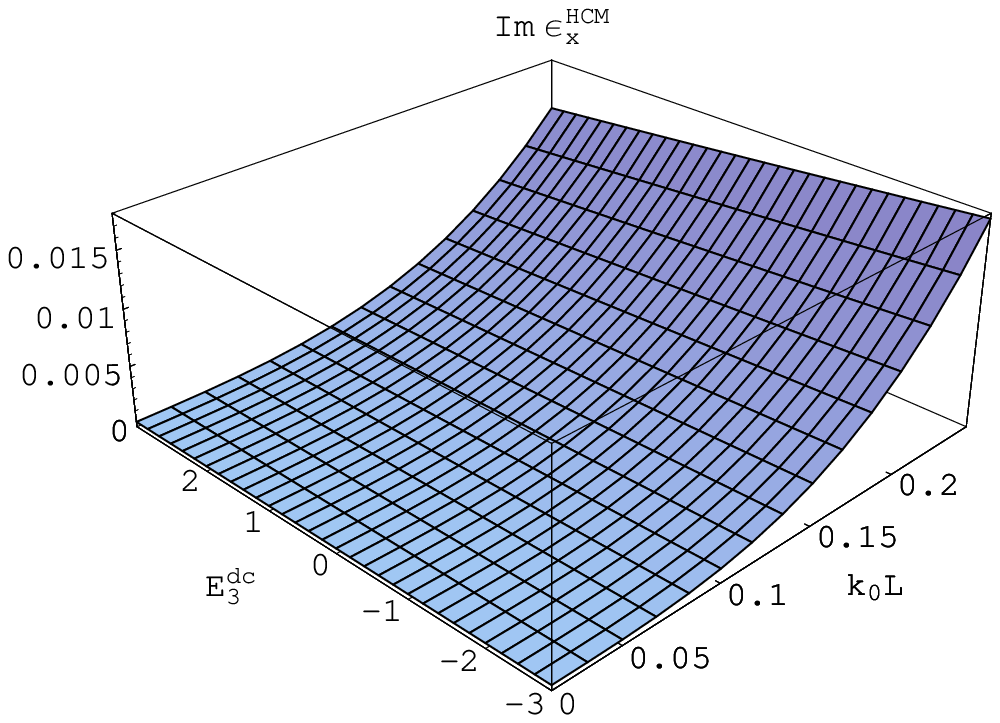,width=3.0in} \vspace{5mm}\\
 \epsfig{file=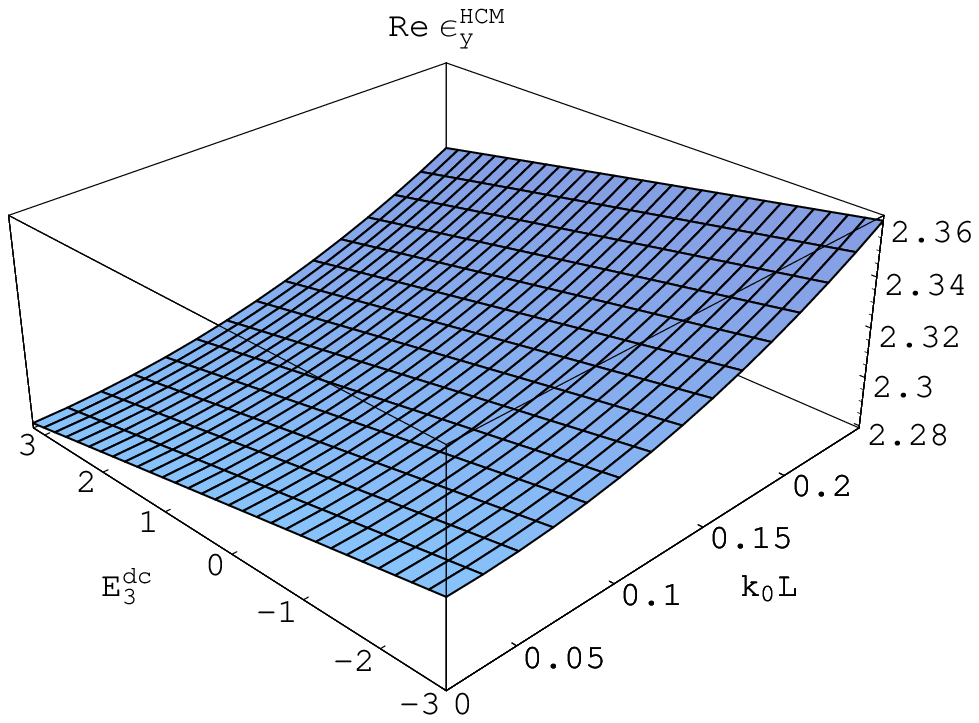,width=3.0in} \hfill
\epsfig{file=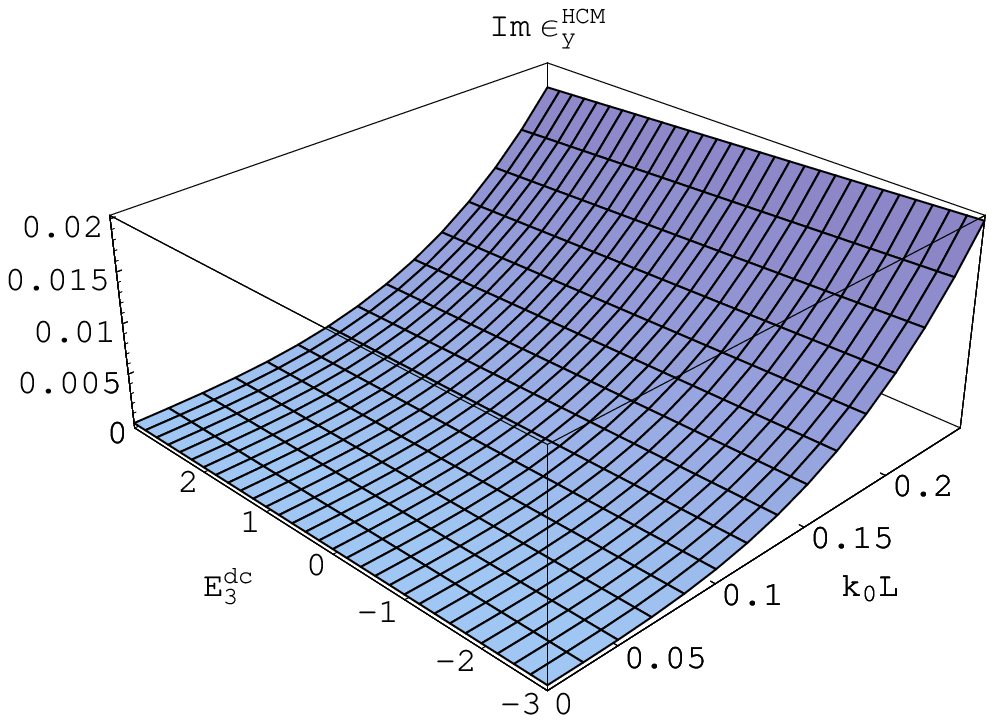,width=3.0in}\vspace{5mm}\\
 \epsfig{file=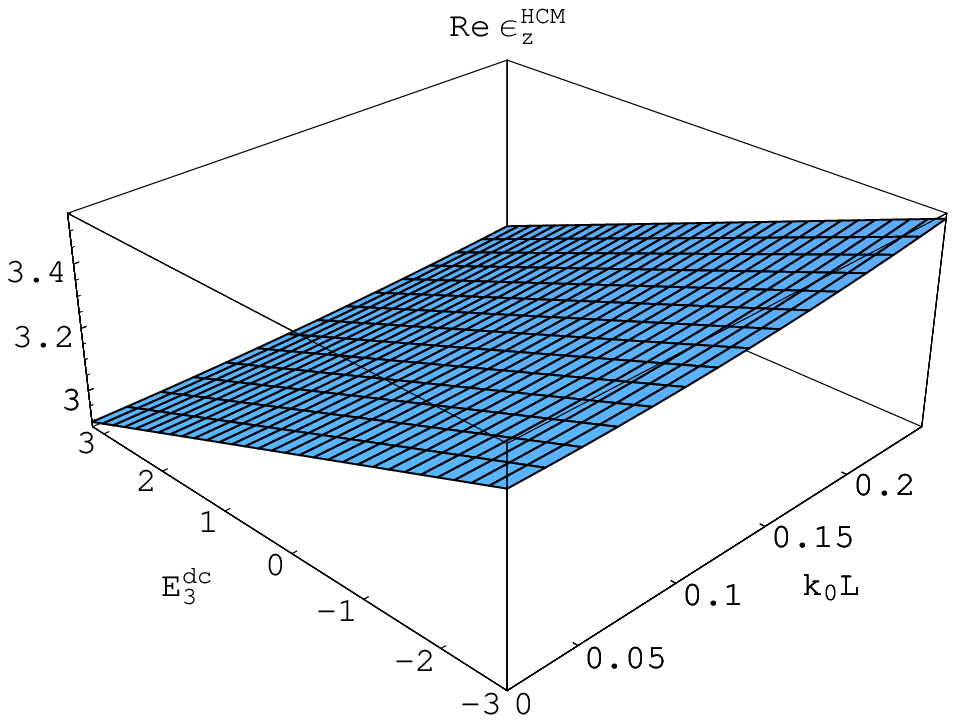,width=3.0in} \hfill
\epsfig{file=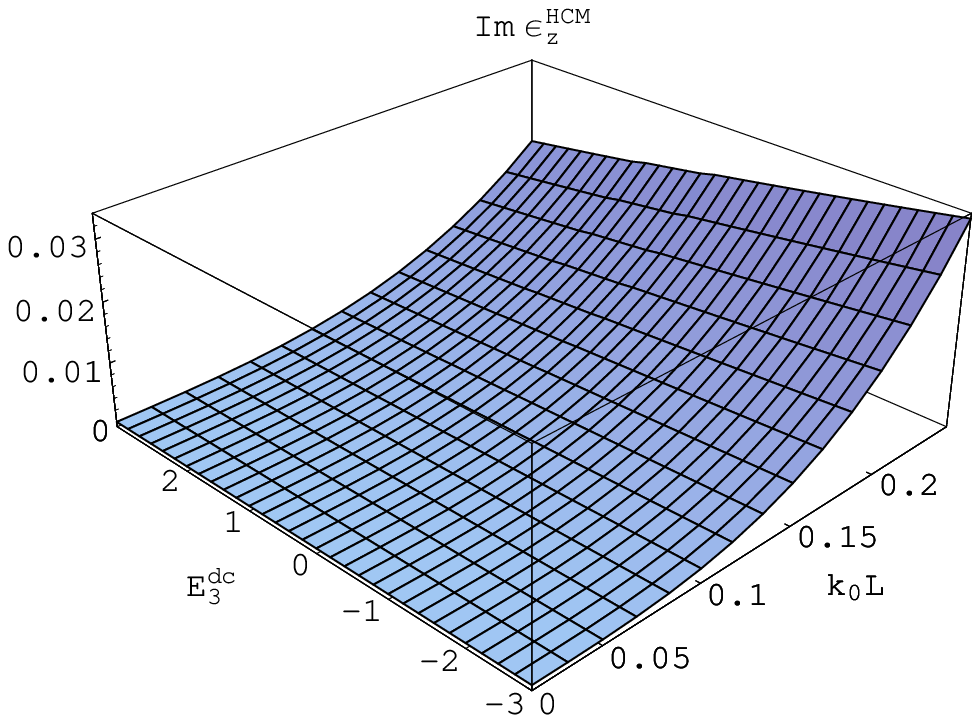,width=3.0in}
  \caption{\label{fig1} Real (left) and imaginary (right)  parts of
  the HCM's relative permittivity scalars $\eps^{HCM}_{x,y,z}$, as estimated
  using the SPFT, plotted against $E^{dc}_{3}$ (in V m${}^{-1}$ $\times
  10^8$) for $E^{dc}_{1,2} = 0$ and $k_{{\scriptscriptstyle 0}} L$.
The constituent material  $a$ is air and the constituent material $b$ is
potassium niobate.
   The particles of both materials $a$ and $b$ are
  parallel prolate spheroids, characterized by
   the shape parameters $\alpha_1 = \alpha_2 =1$
and $\alpha_3 = 9$. Calculations were made in the
limit $\eta\to 0$.
 }
\end{figure}

In relation to the predecessor study \c{LM07}, the effects of two
quantities have to be considered: (i) the correlation length $L$
and (ii) the particle size parameter $\eta$. Let us begin with $L$.
In
Figure~\ref{fig1}, the three relative permittivity scalars  $\eps^{HCM}_{x,y,z}$
computed from \r{SPFT_HCM} in the limit $\eta \to 0$
(by setting $ \=D^{ > 0}=\=0$),  are plotted as
functions of $\ko L$ and $E^{dc}_3 $. At $L = 0$, the variations in
$\eps^{HCM}_{x,y,z}$ as $E^{dc}_3$ increases
 are precisely those
 predicted by the Bruggeman homogenization formalism \c{LM07}; but
the imaginary parts of $\eps^{HCM}_{x,y,z}$
 are nonzero  for $L > 0$. The emergence of these nonzero imaginary parts is
unambiguously  attributable to the incorporation of scattering loss
 in SPFT. As $L$
 increases, the interactions of larger numbers of
constituent particles become correlated, thereby leading to an increase in
the overall scattering loss in the composite material. This is
 most noticeable in the $z$ direction (which is direction of the major principal axis
 of $\=U$ and $\=\eps^{(b)}$, and also the direction of
  $\#E_{dc}$), as indicated in Figure~\ref{fig1} by the behavior of  $\eps^{HCM}_z$.
At $\ko L = 0.2$, the magnitude of the imaginary part of
$\eps^{HCM}_z$ increases by approximately 50\% as $E^{dc}_3$ ranges
from 3 $\times
  10^8$ to $-3$ $\times
  10^8$ V m${}^{-1}$.

\begin{figure}[!ht]
\centering \psfull \epsfig{file=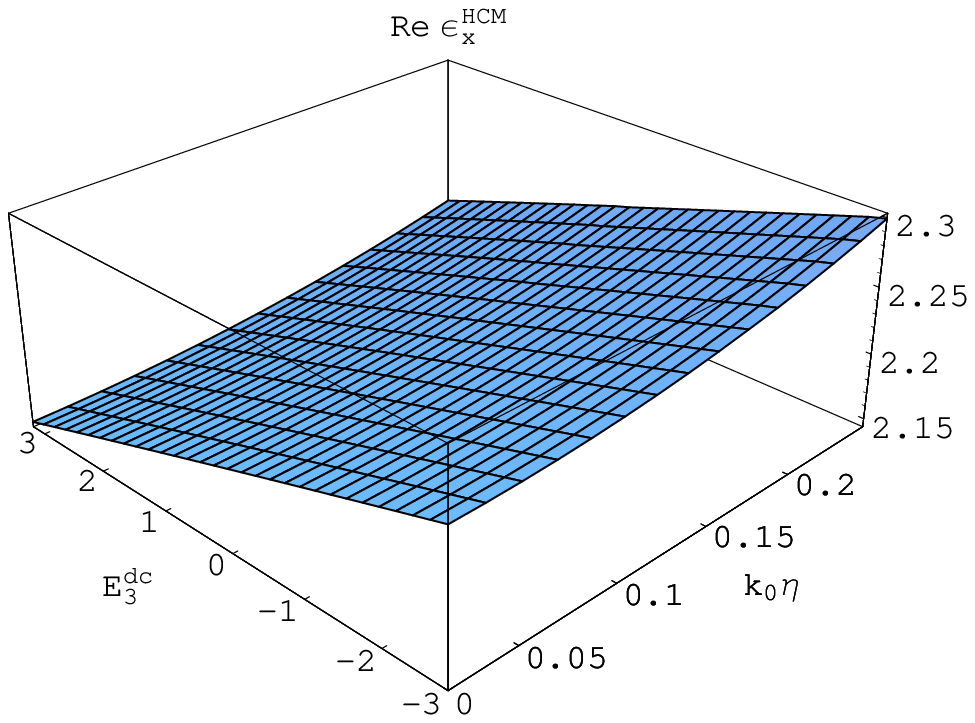,width=3.0in} \hfill
\epsfig{file=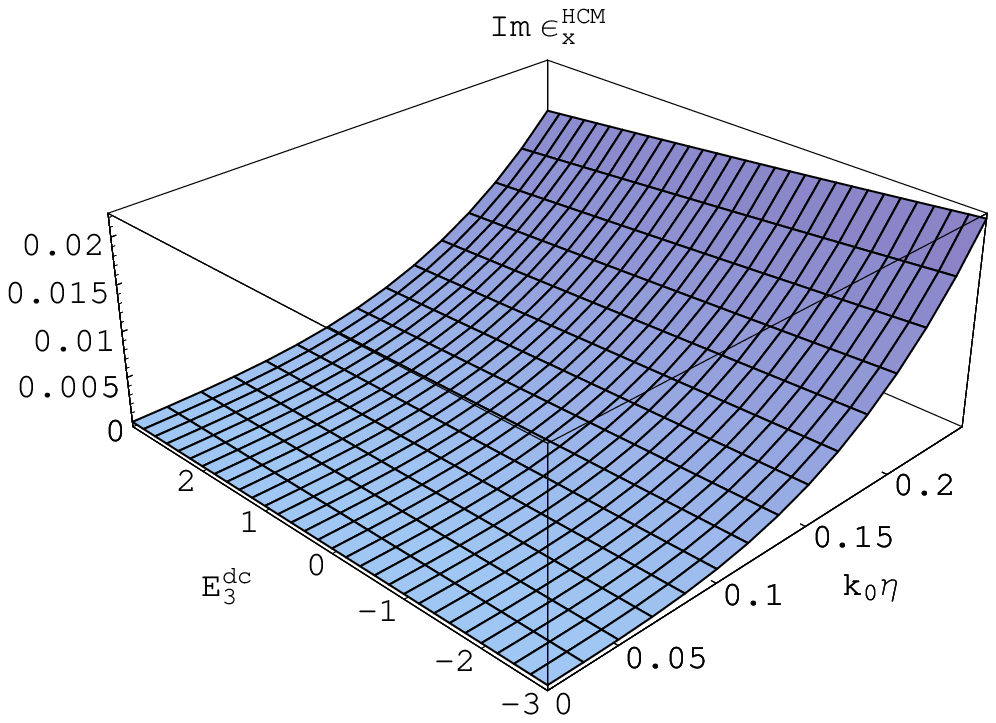,width=3.0in} \vspace{5mm}\\
 \epsfig{file=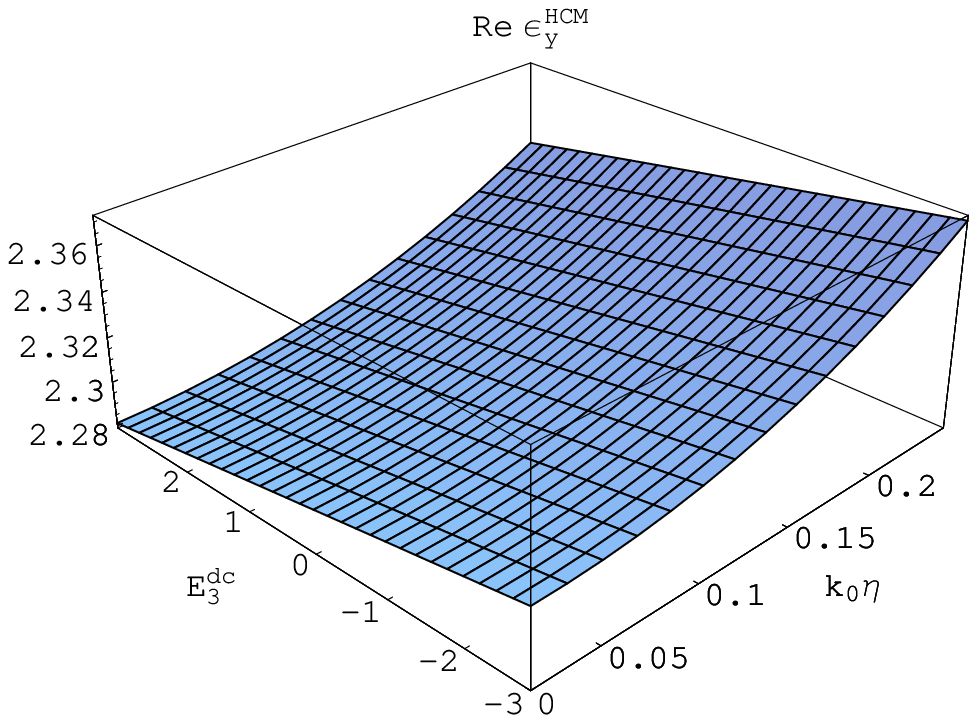,width=3.0in} \hfill
\epsfig{file=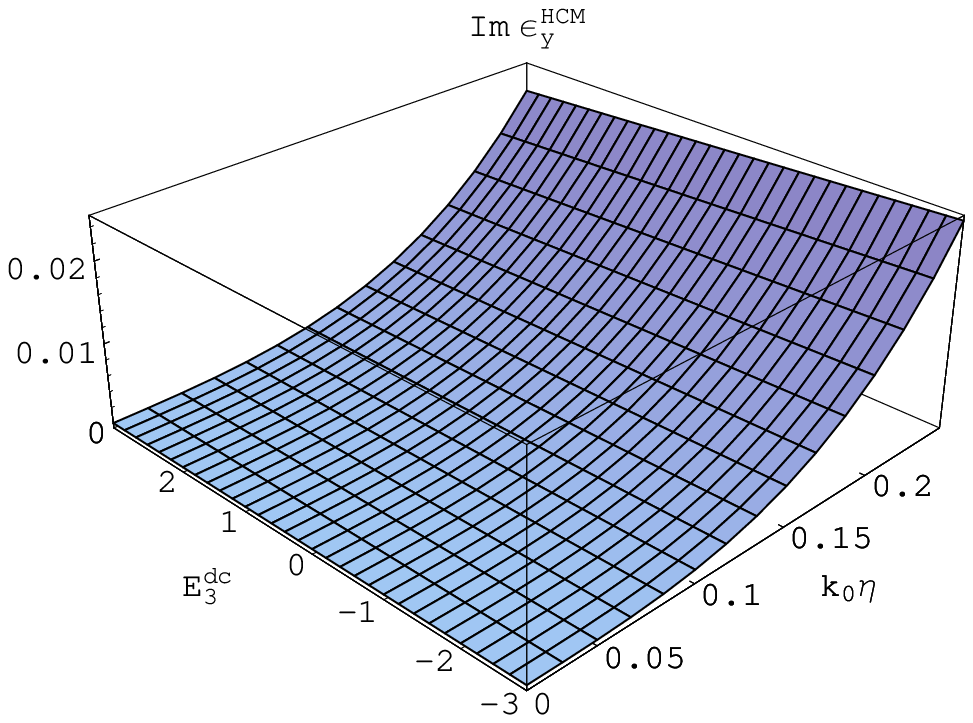,width=3.0in}\vspace{5mm}\\
 \epsfig{file=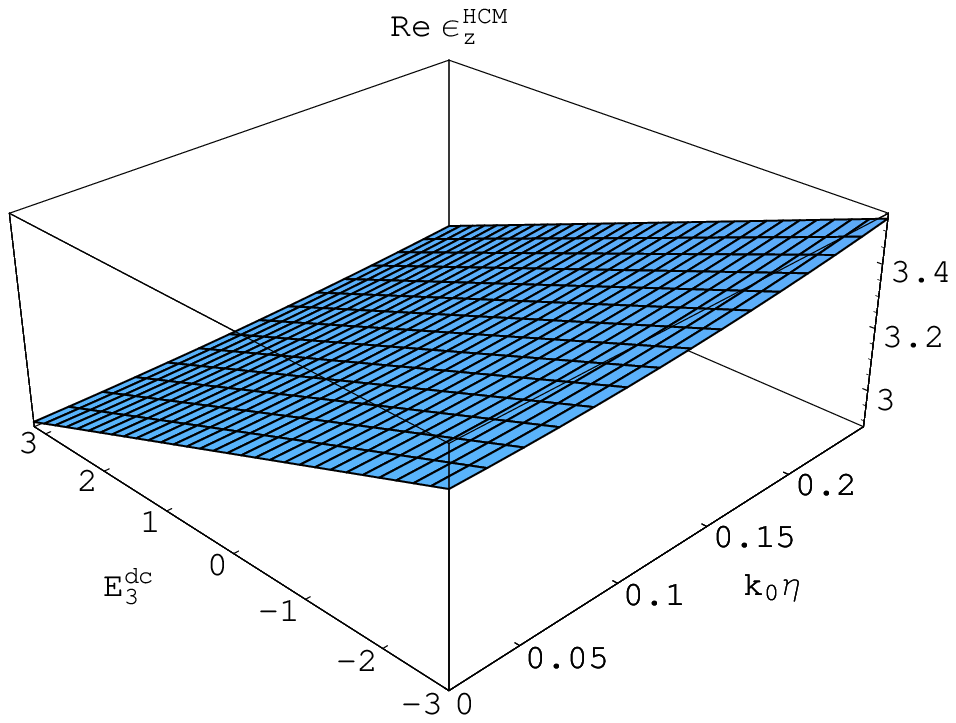,width=3.0in} \hfill
\epsfig{file=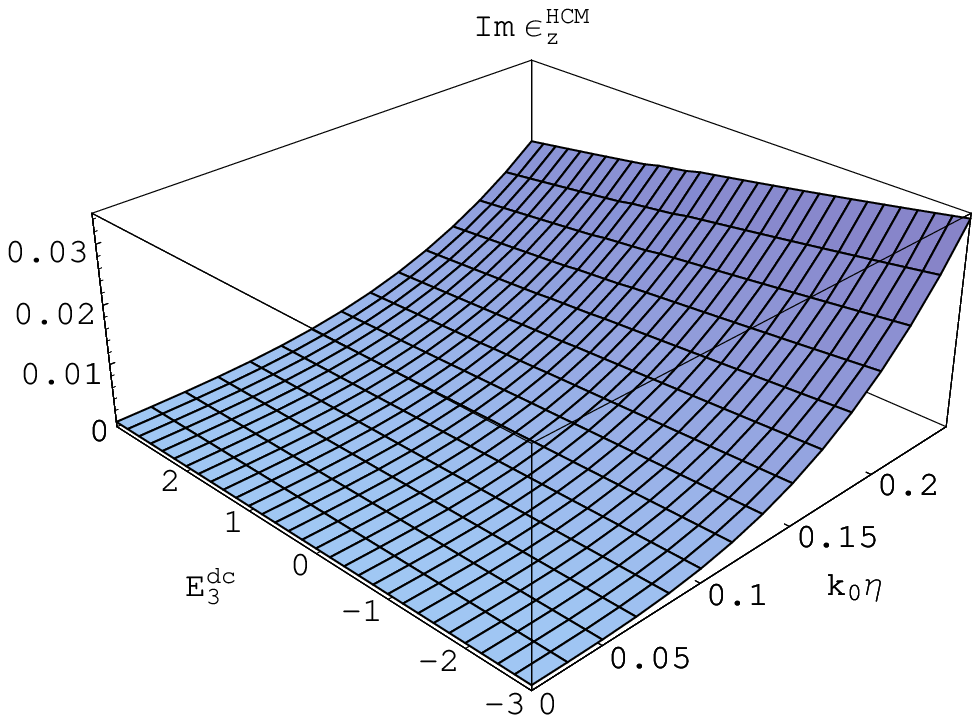,width=3.0in}
  \caption{\label{fig2}
As Figure~\ref{fig1} but with the
  real (left) and imaginary (right)  parts of
  the HCM permittivity scalars $\eps^{HCM}_{x,y,z}$
  plotted against $E^{dc}_{3}$ (in V m${}^{-1}$ $\times
  10^8$)  and $\ko \eta$, with correlation length $L = 0$.}
\end{figure}

The effect of the size parameter $\eta$  is very similar (but not
identical) to that of the correlation length, as is apparent from
Figure~\ref{fig2} wherein the real and imaginary parts of
$\eps^{HCM}_{x,y,z}$ are plotted against $\ko \eta$ and $E^{dc}_3 $,
for $L = 0$. The nonzero imaginary parts of $\eps^{HCM}_{x,y,z}$,
which arise for $\eta > 0$, are also attributable to coherent scattering
losses associated with the finite size of the constituent
particles \c{PLS94}. The effect of the dc electric field over the real and
imaginary parts of the HCM's relative permittivity scalars at a given value
of $\eta $ (with $L =0$) is much the same as it is at the same value
of $L$ (with $\eta =0$).

\begin{figure}[!ht]
\centering \psfull \epsfig{file=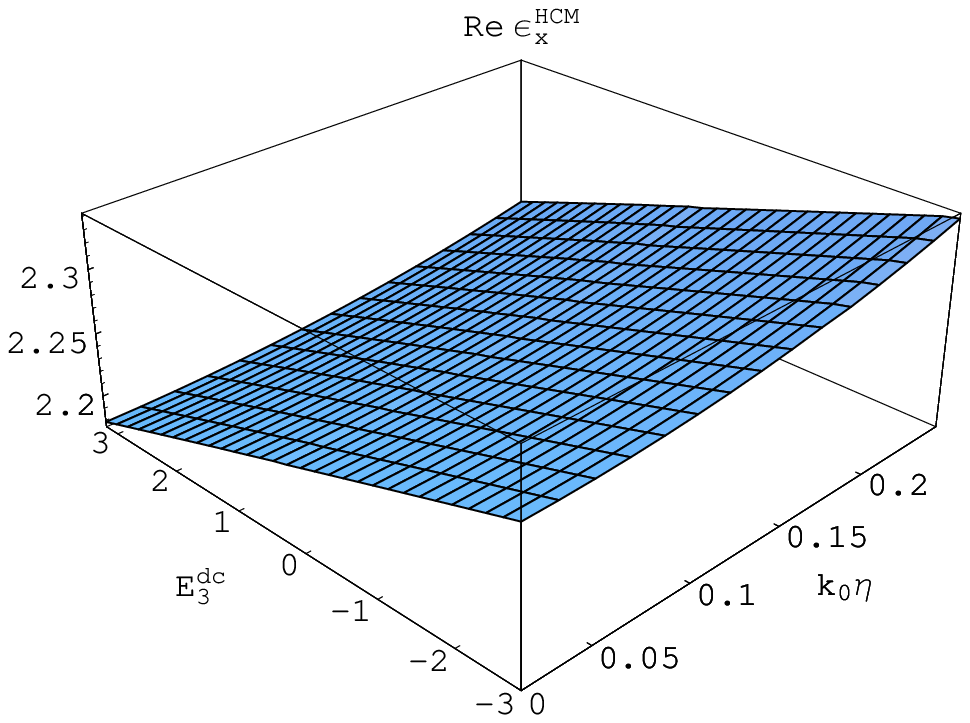,width=3.0in} \hfill
\epsfig{file=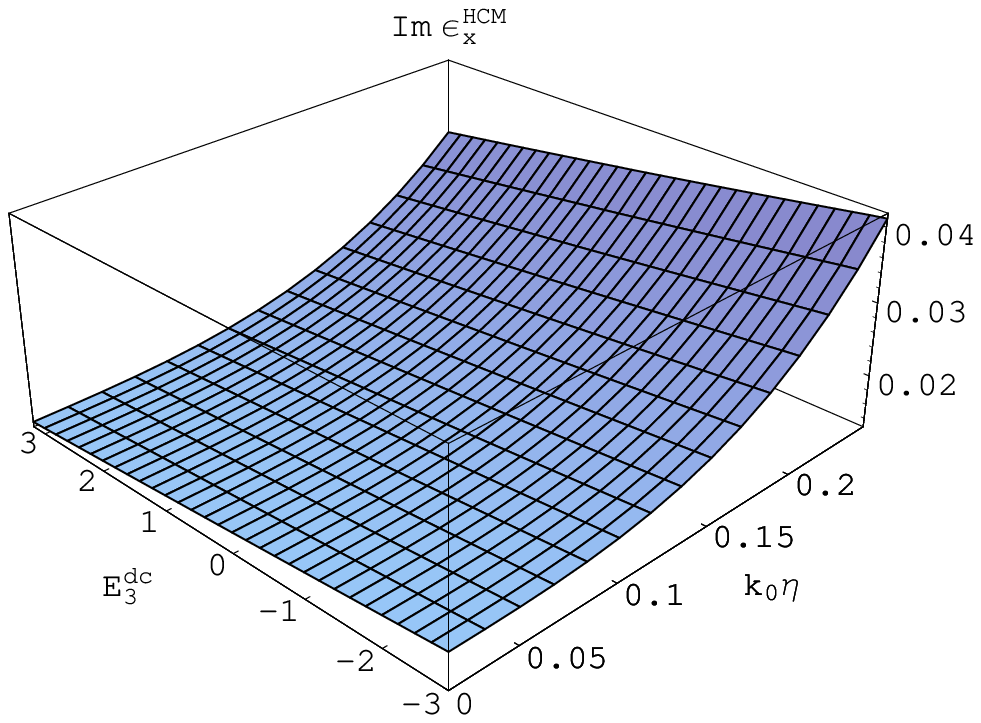,width=3.0in}\vspace{5mm}\\
 \epsfig{file=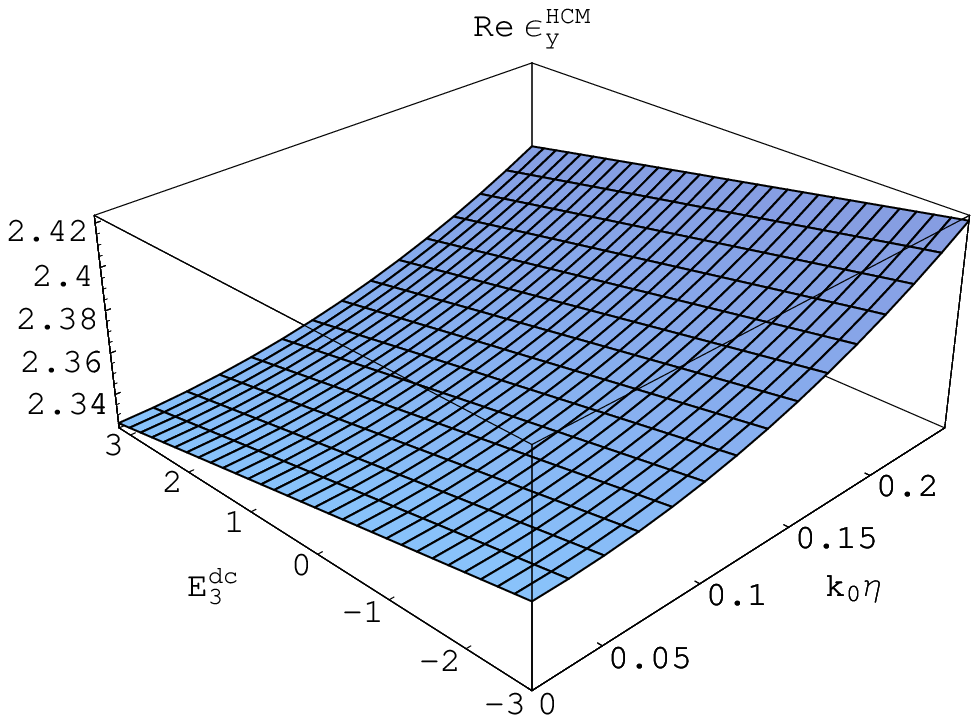,width=3.0in} \hfill
\epsfig{file=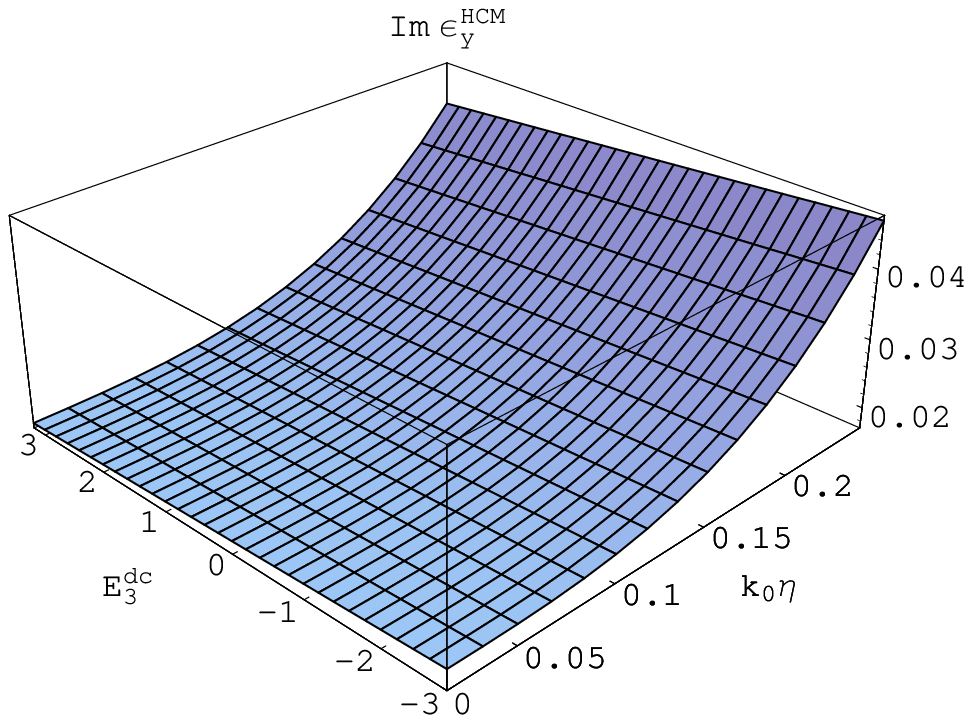,width=3.0in}\vspace{5mm}\\
 \epsfig{file=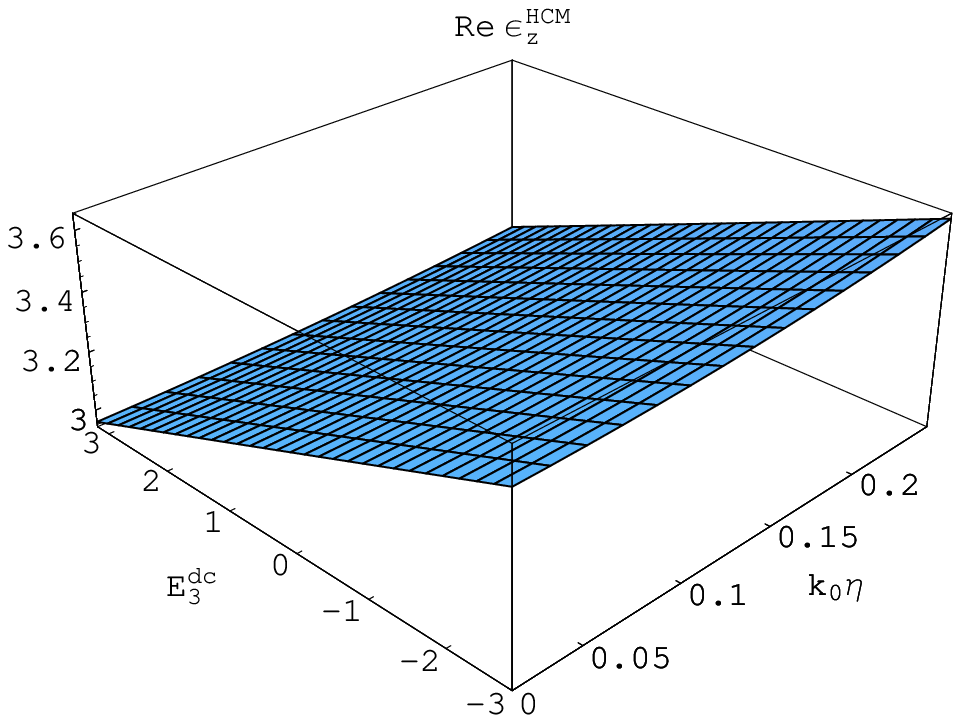,width=3.0in} \hfill
\epsfig{file=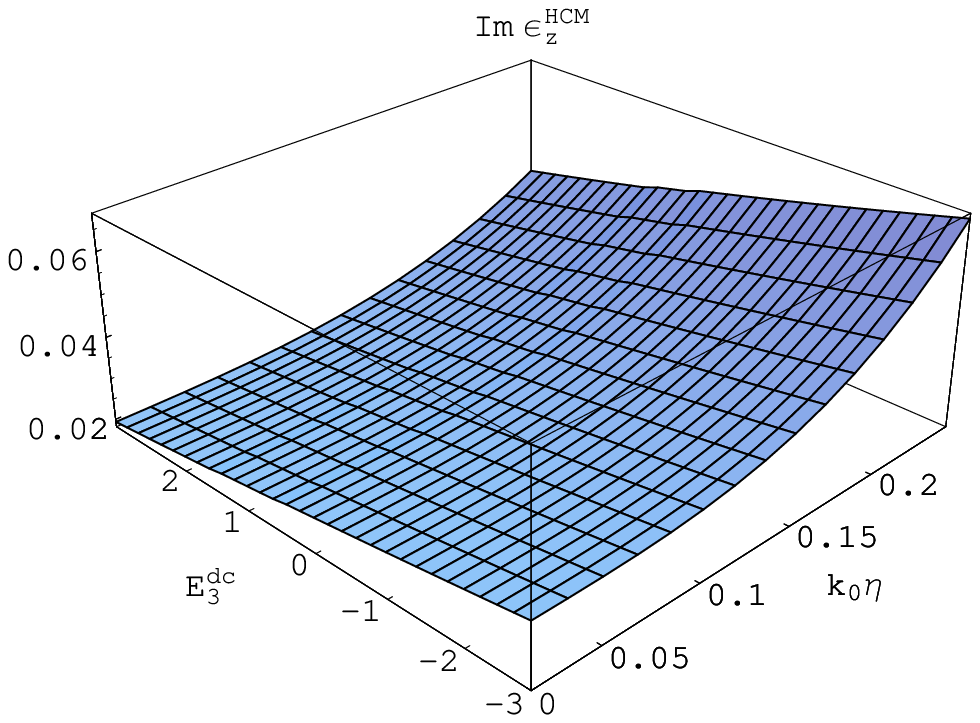,width=3.0in}
  \caption{\label{fig3} As Figure~\ref{fig1} but with the
  real (left) and imaginary (right)  parts of
  the HCM permittivity scalars $\eps^{HCM}_{x,y,z}$
  plotted against $E^{dc}_{3}$ (in V m${}^{-1}$ $\times
  10^8$)  and $\ko \eta$, with  $\ko L = 0.25$.}
\end{figure}

Lastly,  let us turn to the combined influences of the correlation length
and the size of the constituent particles. In Figure~\ref{fig3},  plots of the
real and imaginary parts of $\eps^{HCM}_{x,y,z}$ against $\ko \eta$
and $E^{dc}_3 $ are displayed for the case where  $\ko L = 0.25$.
Plots of the real parts of $\eps^{HCM}_{x,y,z}$ are similar to those
in Figures~\ref{fig1} and \ref{fig2}. In contrast, the magnitudes of
the imaginary parts of $\eps^{HCM}_{x,y,z}$ are substantially larger
in Figure~\ref{fig3} than in Figures~\ref{fig1} and
\ref{fig2}. As is the case in Figure~\ref{fig2}, the imaginary part
of $\eps^{HCM}_{z}$ increases by approximately 50\% as $E^{dc}_3$
ranges from 3 $\times
  10^8$ to $-3$ $\times
  10^8$ V m${}^{-1}$, at $\ko \eta = 0.2$.

Let us conclude by stating that
a particular feature of the electromagnetic response of an HCM is
that
  attenuation can arise, due to coherent scattering loss, regardless of
  whether
  the  constituent materials are dissipative or nondissipative.
  The extended SPFT \c{M04} provides
a means of estimating the effect of this scattering loss on the
effective constitutive parameters of the HCM, and relating it to the
distributional statistics of the particles which make up the
constituent materials and their size. When one (or more) of the
constituent materials exhibits the Pockels effect, the degree of
scattering loss may be significantly controlled by the application
of a low--frequency (dc) electric field. The technological
implications of this capacity to control dynamically, at the
post--fabrication stage,  the electromagnetic properties of an
HCM~---~which itself may be tailored to a high degree at the
fabrication stage~---~are far--reaching, extending to applications
in telecommunications, sensing, and actuation, for example.

\vspace{10mm}

\noindent{\bf Acknowledgement:} TGM is supported by a \emph{Royal
Society of Edinburgh/Scottish Executive Support Research
Fellowship}.\\

\newpage

\newpage

\newpage

\end{document}